\documentclass[twocolumn,prl,amsmath,amssymb,showpacs,superscriptaddress]{revtex4}
\usepackage{graphicx}
\pdfoutput=1

\begin{document}

\title{Aharonov-Casher Effect in Wigner Crystal Exchange Interactions}

\author{Yaroslav Tserkovnyak}
\affiliation{Department of Physics and Astronomy, University of California, Los Angeles, California 90095, USA}
\author{Markus Kindermann}
\affiliation{School of Physics, Georgia Institute of Technology, Atlanta, Georgia 30332, USA}

\date{\today}

\begin{abstract}
We theoretically study the effects of spin-orbit coupling on spin exchange in a low-density Wigner crystal.  In addition to the familiar antiferromagnetic Heisenberg exchange, we find general anisotropic interactions in spin space if the exchange paths allowed by the crystal structure form loops in real space. In particular, it is shown that the two-electron exchange interaction can acquire ferromagnetic character.
\end{abstract}



\maketitle

To a first approximation, the electrons in a Wigner crystal are localized in space through their mutual Coulomb repulsion.  The crystal electrons  are thus distinguishable through their location on the crystal lattice.  For any finite interaction strength, however, the electrons are able to tunnel through the localizing Coulomb barrier, loosing this distinguishability. For spinless particles, such tunneling is largely inconsequential, since the  electron configurations before and after tunneling are equivalent. Because they carry spin, however, a  process in which two crystal electrons exchange their positions does have an effect: it exchanges the spins on the respective lattice sites. While the charge configuration of the crystal is still nearly static, its spins acquire some dynamics.

In the absence of spin-orbit interaction (SOI), the corresponding two-electron exchange interaction takes the Heisenberg form, $J\mathbf{S}_1\cdot\mathbf{S}_2$, by spin-rotational symmetry. Because typically the lowest-energy orbital wave functions are  symmetric under particle interchange, the spin exchange is generically antiferromagnetic ($J>0$), as a consequence of the Pauli principle. This statement has  been established rigorously for one-dimensional (1D) many-electron systems with velocity- and spin-independent interactions \cite{liebPR62}. The SOI, however, has long been known to modify the Heisenberg form of the exchange Hamiltonian by, e.g., effectively canting the participating spins. The resulting Dzyaloshinsky-Moriya (DM) interaction \cite{dzyaloshinskyJPCS58} of the form $\mathbf{D}\cdot(\mathbf{S}_1\times\mathbf{S}_2)$, with some structure-dependent  vector $\mathbf{D}$, was initially studied in the context of weak ferromagnetism in materials that otherwise were expected to be antiferromagnetic, but has since been appreciated in many other contexts.

\begin{figure}
\centerline{\includegraphics[width=0.8\linewidth]{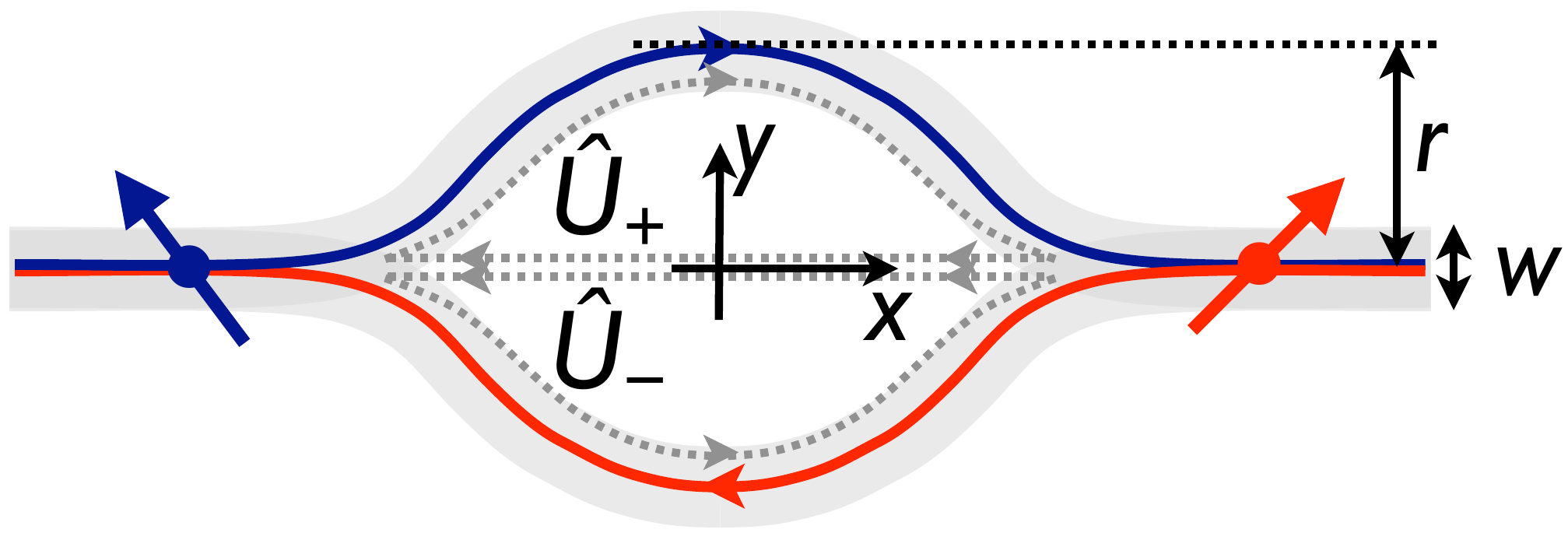}}
\caption{(Color online) Electrons confined to move along the $x$ axis undergo a geometric spin transformation as they exchange their positions. The shaded area marks the transverse confinement of width $w$. As the electrons tunnel through the mutual Coulomb repulsion barrier, they deviate laterally by the distance $r\sim w(w/a_B)^{1/3}$, in the center-of-mass reference frame shown here. Sketched are the classical trajectories following the rim of the inverted potential. If the left electron chooses to follow the upper path, the right electron is forced to simultaneously go along the lower path and vice versa. $\hat{U}_\pm$ is the SOI SU(2) transformation (\ref{tU}) starting at the origin of the $xy$ coordinate system, moving to the left, then (counter)clockwise along the upper (lower) rim and back. The total area enclosed by the two trajectories is $A\sim r^2$.}
\label{fig}
\end{figure}

This paper continues the recent discussion of the SOI in mesoscopic and Wigner-crystal exchange processes \cite{kavokinPRB01,stepanenkoPRB03,imamuraPRB04, gangadharaiahPRL08}. For definiteness, we focus our attention on clean strongly interacting quasi-1D systems \cite{auslaenderSCI05}. Two-electron as well as ring exchange interactions in single- and double-row quasi-1D systems without SOI were extensively discussed in Refs.~\cite{matveevPRL04,klironomosPRB05}. We  show that the SOI qualitatively enriches the electronic behavior in 1D by causing anisotropies in the exchange Hamiltonian.  Our calculation is performed using the  path-integral instanton picture of particle exchange \cite{voelkerPRB01,klironomosPRB05}. In the low-density limit, $a_B\ll b$, where $b=n^{-1}$ is the average inter-electron separation and $a_B=\varepsilon\hbar^2/me^2$ is the Bohr radius, the dominant electron paths   follow classical trajectories in the inverted potential. The SOI is then naturally captured by purely geometric SU(2) spin transformations along the electronic exchange paths \cite{tserkovPRB07ac}. See Fig.~\ref{fig} for a schematic. We thus explore  the non-Abelian Aharonov-Casher \cite{aharonovPRL84} variant of physics whose Aharonov-Bohm counterpart was studied extensively in Refs.~\cite{okamotoPRB98}.

To illustrate our approach and the key findings, we consider two spin-$1/2$ electrons confined to move in the $xy$ plane, subject to the two-dimensional (2D) effective-mass Hamiltonian (denoting by hats the spin structure):
\begin{align}
\hat{H}=&\sum_{j=1,2}[(-i\hbar\boldsymbol{\nabla}_j+\hat{\mathbf{A}}_j)^2/2m+V_j]+U_{12}\,.
\label{H}
\end{align}
Here, $\boldsymbol{\nabla}_j=(\partial_{x_j},\partial_{y_j})$ and $\mathbf{A}_j=\hat{\mathbf{A}}(\mathbf{r}_j)=(\hat{A}_x,\hat{A}_y)$ is the two-component $2\times2$ vector potential, which can be expanded in the basis of the three Pauli matrices $(\hat{\sigma}_x,\hat{\sigma}_y,\hat{\sigma}_z)\equiv\hat{\boldsymbol{\sigma}}$. The SU(2) vector potential $\hat{\mathbf{A}}$ describes the SOI. The Pauli equation for electrons in vacuum, for example, has $\hat{\mathbf{A}}\propto\boldsymbol{\nabla}V\times\hat{\boldsymbol{\sigma}}$, while in 2D electron systems with asymmetric confining potential along the $z$ axis, the SOI is usually dominated by the so-called Rashba  interaction of the form: $\hat{\mathbf{A}}=\alpha\,\mathbf{z}\times\hat{\boldsymbol{\sigma}}$. Further, $U_{12}=U(\mathbf{r}_1-\mathbf{r}_2)$ is the effective two-particle repulsion and $V_j=V(\mathbf{r}_j)$ is the 2D confinement potential along the $x$ axis. The two-electron exchange then becomes a quasi-1D scattering problem: In the center-of-mass reference frame, electrons are incident from  opposite directions and are scattering near the origin, $\Delta\mathbf{r}=\mathbf{r}_1-\mathbf{r}_2\sim0$. The electron transmission through the repulsion potential $U(\Delta\mathbf{r})$ in the low-density limit is well approximated by  WKB tunneling.

We  start by assuming a homogeneous Rashba SOI. In the strictly 1D limit, when the lateral motion along the $y$ direction can be completely neglected, it is possible to gauge out the SOI by the  unitary transformation
\begin{equation}
\hat{W}(x)=e^{-ix\hat{A}_x/\hbar}\,.
\label{g1D}
\end{equation}
When applied  to each electron separately, this gauge transformation removes   the vector potential $\hat{\mathbf{A}}$ from the Hamiltonian. It is, however, not possible to generalize this to 2D, since $\hat{A}_x$ and $\hat{A}_y$ in general do not commute. We, nevertheless, choose to transform our 2D Hamiltonian (\ref{H}) as $\hat{H}\to\hat{W}^\dagger\hat{H}\hat{W}$. When the electrons are sufficiently distant from each other and they stay close to $y=0$ due to the $V(y)$ confinement,   the transformation (\ref{g1D}) then does remove the SOI from the problem. Only during the short time intervals when the electrons tunnel through each other do they deviate appreciably from  the $x$ axis and undergo an additional SOI-induced spin transformation (see Fig.~\ref{fig}). It is this remnant spin precession that the gauge (\ref{g1D}) allows us to focus on. To simplify our discussion, we  assume that the spin precession length $l_{\rm so}=\pi\hbar/\alpha$ is much longer than the width $w=\sqrt{\hbar/m\omega}$  of the wire (assuming quadratic lateral confinement $V=m\omega^2y^2/2$), while the Bohr radius $a_B$ is much shorter than $w$: $a_B\ll w\ll l_{\rm so}$. In the opposite limit, $w\ll a_B$, electrons are confined very tightly in the lateral direction, forcing them to tunnel ``head on" \cite{klironomosPRB05} and thus invalidating the simple instanton calculation of the geometric SOI phase. In the case of 2D and 3D Wigner crystals, it would be sufficient to require that $a_B\ll b$, which guarantees that the electrons avoid head-on tunneling. In the case of a quasi-1D model, it is convenient (although not essential) to assume $b\gg r\sim w(w/a_B)^{1/3}$, which allows one to neglect the tendency of 1D crystals to form zigzag arrangements \cite{klironomosPRB05}.

Before proceeding, it is instructive to recall how one constructs the exchange Hamiltonian in the low-density limit without   SOI using the instanton method. The strength of the exchange is parametrized by a positive real-valued parameter $J$, which is determined by the Euclidian action $S_E$ along the minimal-action path exchanging two electrons \cite{voelkerPRB01,klironomosPRB05}: $J=\beta\hbar\omega_0\sqrt{S_E/2\pi\hbar}\,e^{-S_E/\hbar}$, where $\omega_0$ is the characteristic attempt frequency (corresponding to the effective electron confinement along the wire) and $\beta$ is a numerical prefactor of order unity. The classical minimal action path corresponds approximately to the rim of the inverted potential $U_E(\Delta\mathbf{r})=-[V(\Delta y)/2+U(\Delta\mathbf{r})]$ (again assuming quadratic confinement) in the center-of-mass frame: $S_E\approx\int_{-b}^{b} dl\sqrt{2m^\ast |U_E(l)|}$. $J$ is thus essentially the WKB tunneling amplitude for a particle with the reduced mass $m^\ast=m/2$ through the potential barrier $U_E$ along the classical trajectory parametrized by $l$. The   authors of Ref.~\cite{klironomosPRB05}  performed a detailed study of the Euclidian action in quasi-1D wires, obtaining in the regime of $a_B\ll w,r\ll b$: $S_E/\hbar\approx\eta\sqrt{b/a_B}-\kappa(w/a_B)^{2/3}$, where $\eta$ and $\kappa$ are numerical prefactors of order unity. The first term on the right-hand side is the action for the head-on tunneling of electrons and the second term is its reduction due to lateral excursions that allow electrons to  avoid each other, as sketched in Fig.~\ref{fig}.

In the absence of the SOI, $J$ parametrizes the usual antiferromagnetic coupling
 between the two spins \cite{klironomosPRB05}:
\begin{equation}
\hat{H}_{\rm AF}=J(\hat{1}\otimes\hat{1}+\hat{\boldsymbol{\sigma}}\otimes\hat{\boldsymbol{\sigma}})\,,
\label{HAF}
\end{equation}
making the convention that the left (right) operator in a tensor product $\hat{L}\otimes\hat{R}$ acts on the left (right) spin. The operator $\hat{X}=(\hat{1}\otimes\hat{1}+\hat{\boldsymbol{\sigma}}\otimes\hat{\boldsymbol{\sigma}})/2$ simply exchanges the spins: $\hat{X}|s_1,s_2\rangle=|s_2,s_1\rangle$, where $|s_1,s_2\rangle$ is the spinor wave function with $s_1$ ($s_2$) corresponding to the left (right) spin, respectively. In the case of multiple exchange trajectories, $J$ stands for the sum of all contributions. We now include SOI effects. During the position exchange, each electron that is moving along path $\mathcal{C}$ undergoes the path-dependent SU(2) transformation \cite{tserkovPRB07ac}
\begin{equation}
\hat{U}=T_{\mathcal{C}}\exp^{-i\oint_{\mathcal{C}}d\mathbf{r}\cdot\hat{\mathbf{A}}/\hbar}\,,
\label{tU}
\end{equation}
where $T_\mathcal{C}$ is the contour-ordering operator and the integral runs over the classical exchange paths closed along the $x$ axis, as shown in Fig.~\ref{fig}. We choose the point where the two electrons pass each other to be  at $x=0$, see Fig.~\ref{fig}, and integrate (counter)clockwise along the upper (lower) classical exchange trajectory. The SOI is assumed to be weak in comparison to the interparticle repulsion, such that the effect of the spin dynamics on the instanton orbital trajectories may be neglected.

The effective spin Hamiltonian is finally obtained as:
\begin{equation}
\hat{H}_s=J(\hat{U}_+^\dagger\otimes\hat{U}_-+\hat{U}_-^\dagger\otimes\hat{U}_+)(\hat{1}\otimes\hat{1}+\hat{\boldsymbol{\sigma}}\otimes\hat{\boldsymbol{\sigma}})/2\,.
\label{HS}
\end{equation}
The operator $(\hat{U}^\dagger_-\otimes\hat{U}_+)\hat{X}$ implements  the spin transformation along the clockwise exchange trajectory, when the left electron deviates from $y=0$ toward positive $y$, moving to the right, while the right electron is correspondingly pushed toward negative $y$, moving to the left. The other term  $\hat{U}^\dagger_+\otimes\hat{U}_-$ accounts for the counterclockwise exchange path. $\hat{H}_s$ is verified to be  Hermitian through the identity $\hat{X}(\hat{U}_+\otimes\hat{U}_-^\dagger)=(\hat{U}^\dagger_-\otimes\hat{U}_+)\hat{X}$. Notice that electrons whose exchange occurs at $x=x_0$ instead of $x=0$, as in Fig.~\ref{fig}, acquire an additional rotation of their interaction Hamiltonian by $\hat{W}(x_0)$: $\hat{H}_s\to\hat{W}^\dagger(x_0)\hat{H}_s\hat{W}(x_0)$, with $\hat{W}$ applied to both spins.
 
We now employ   Eq.~(\ref{HS}) to calculate the spin exchange coupling in the presence of the SOI, Eq.~(\ref{H}). Parametrizing $\hat{U}_\pm=u_\pm-i\mathbf{u}_\pm\cdot\hat{\boldsymbol{\sigma}}$ with real-valued scalars $u_\pm$ and vectors $\mathbf{u}_\pm$, such that $u_\pm^2+\mathbf{u}_\pm^2=1$, the Hamiltonian (\ref{HS})   acquires the   form (omitting a constant offset):
\begin{align}
\hat{H}_s/J=&(u_+u_--\mathbf{u}_+\cdot\mathbf{u}_-)\hat{\boldsymbol{\sigma}}_1\cdot\hat{\boldsymbol{\sigma}}_2+\mathbf{d}\cdot(\hat{\boldsymbol{\sigma}}_1\times\hat{\boldsymbol{\sigma}}_2)\nonumber\\
&+2(\mathbf{a}\cdot\hat{\boldsymbol{\sigma}}_1)(\mathbf{a}\cdot\hat{\boldsymbol{\sigma}}_2)-2(\mathbf{b}\cdot\hat{\boldsymbol{\sigma}}_1)(\mathbf{b}\cdot\hat{\boldsymbol{\sigma}}_2)\,,
\label{HS2}
\end{align}
where $\mathbf{d}=u_+\mathbf{u}_-+u_-\mathbf{u}_+$ parametrizes DM, $\mathbf{a}=(\mathbf{u}_++\mathbf{u}_-)/2$ Ising antiferromagnetic, and $\mathbf{b}=(\mathbf{u}_+-\mathbf{u}_-)/2$ Ising ferromagnetic interactions. The operators $\hat{\boldsymbol{\sigma}}_1$  and $\hat{\boldsymbol{\sigma}}_2$ act on the left and right spin, respectively.

Specializing to the Rashba case and expanding the transformation matrices in $\gamma=\sqrt{A}(\alpha/\hbar)$ (where $A$ is the total area of the loop formed by the two classical trajectories in Fig.~\ref{fig}), we have \cite{tserkovPRB07ac}:
\begin{equation}
\hat{U}_\pm=1\mp i\gamma^2\hat{\sigma}_z+i\gamma^3\mathbf{v}_\pm\cdot\hat{\boldsymbol{\sigma}}+\mathcal{O}(\gamma^4)\,.
\label{Upm}
\end{equation}
Since the cubic-order in $\gamma$ contribution to $\hat{U}_\pm$ depends on the exact shape of the exchange loop, we parametrized it by the dimensionless spin-space vectors $\mathbf{v}_\pm$. The same is true also of the higher-order terms. Substituting the expansion (\ref{Upm}) into Eq.~(\ref{HS2}) gives:
\begin{equation}
\hat{H}^{\rm pert}_s/J=\hat{\boldsymbol{\sigma}}_1\cdot\hat{\boldsymbol{\sigma}}_2-\gamma^3\mathbf{v}\cdot(\hat{\boldsymbol{\sigma}}_1\times\hat{\boldsymbol{\sigma}}_2)-2\gamma^4\hat{\sigma}_{1z}\hat{\sigma}_{2z}\,,
\label{HSa}
\end{equation}
where $\mathbf{v}=\mathbf{v}_++\mathbf{v}_-$. In addition to the leading antiferromagnetic coupling, we  find a DM interaction \cite{dzyaloshinskyJPCS58} at order $\gamma^3$ and a ferromagnetic Ising coupling along the $z$ axis at   ${\cal O}(\gamma^4)$. In the approximation (\ref{HSa}), we retained only the leading in $\gamma$ terms separately for Heisenberg, DM, and Ising interactions. If the wire is mirror symmetric with respect to the $xz$ plane, we have $u_+=u_-$ and $\mathbf{u}_-=-\hat{M}\mathbf{u}_+$, where $\hat{M}$ stands for the mirror image, so that $\mathbf{u}_++\mathbf{u}_-=\mathbf{u}_+-\hat{M}\mathbf{u}_+\propto\mathbf{y}$. This means, in particular, that $\mathbf{d}\propto\mathbf{y}$ in Eq.~(\ref{HS2}) and the DM interaction $\hat{H}_{\rm DM}\propto\mathbf{y}\cdot(\hat{\boldsymbol{\sigma}}_1\times\hat{\boldsymbol{\sigma}}_2)$ is of the general form allowed by the mirror symmetry, to all orders in $\gamma$. Furthermore, a  Rashba system is mirror symmetric with respect to the $xy$ plane (combined with flipping $\alpha\to-\alpha$). This constrains the DM coupling  to be odd and the Heisenberg and Ising terms even in $\alpha$, in accordance with Eq.~(\ref{HSa}).

A ferromagnetic Ising coupling of the form $\hat{\sigma}_{1z}\hat{\sigma}_{2z}$ is also expected as a consequence of correlated orbital quantum fluctuations, which produce  van der Waals-type spin interactions via SOI   \cite{gangadharaiahPRL08}.  A form of the exchange similar to our Eq.~(\ref{HSa}), consisting of the Heisenberg, Ising, and DM pieces has been reported before in Refs.~\cite{kavokinPRB01,stepanenkoPRB03,gangadharaiahPRL08}. An analogous result was also found for the RKKY interaction mediated by itinerant electrons in the presence of the Rashba SOI \cite{imamuraPRB04}. In contrast to our Eq.~(\ref{HS2}), however, Refs.~\cite{gangadharaiahPRL08,kavokinPRB01,stepanenkoPRB03,imamuraPRB04} predicted a spin exchange of the form:
\begin{equation} \label{Hs0}
\hat{H}^{(0)}_s=J\hat{\boldsymbol{\sigma}}_1\cdot\hat{\boldsymbol{\sigma}}_2+\Upsilon\mathbf{n}\cdot(\hat{\boldsymbol{\sigma}}_1\times\hat{\boldsymbol{\sigma}}_2)+\Gamma(\mathbf{n}\cdot\hat{\boldsymbol{\sigma}}_1)(\mathbf{n}\cdot\hat{\boldsymbol{\sigma}}_2)\,,
\end{equation}
 parametrized by a single vector $\mathbf{n}$. This stemmed from a hidden SU(2) symmetry of the exchange Hamiltonian in Refs.~\cite{kavokinPRB01,imamuraPRB04} (when the Hamiltonian can be written as a Heisenberg exchange between canted spins) and from a  spin-rotational symmetry around $\mathbf{n}$ in Ref.~\cite{stepanenkoPRB03}. Such symmetries are not assumed in our calculation based on Eq.~(\ref{HS}), allowing for the more general Hamiltonian (\ref{HS2}).

To elucidate the anisotropic spin structure of the exchange Hamiltonian (\ref{HS2}), we distinguish two effects of the SOI. First, the SO coupling  cants the participating spins through a spin rotation along the exchange path. We have already gauged out the main part of that rotation by means of our transformation $\hat{W}$, but deviations of the exchange paths  from the $x$ axis add another piece, contributing to $\hat{U}_\pm$. A canting of  two spins that participate in the usual Heisenberg exchange  \cite{kavokinPRB01,imamuraPRB04,gangadharaiahPRL08} by a rotation angle $\theta$ around the direction $\mathbf{n}$, $\hat{H}_{\mathbf{n}}/J=(\hat{V}_1\hat{\boldsymbol{\sigma}}_1\hat{V}^\dagger_1)\cdot(\hat{V}_2^\dagger\hat{\boldsymbol{\sigma}}_2\hat{V}_2)$,
where $\hat{V}_j=e^{-i\mathbf{n}\cdot\hat{\boldsymbol{\sigma}}_j\,\theta/2}$, produces an exchange Hamiltonian of the form (\ref{Hs0}), with $\Upsilon=-J\sin2\theta$ and $\Gamma=J(1-\cos\theta)$. $\hat{H}_{\mathbf{n}}$, however, has the same eigenvalues as the isotropic Heisenberg exchange Hamiltonian (\ref{HAF}) \cite{shekhtmanPRL92,gangadharaiahPRL08}: A mere canting of two spins should not be considered a real anisotropy, although the resulting Hamiltonian has a DM and an Ising pieces.

Our exchange geometry has a  2D character with two different exchange paths corresponding to transformations $\hat{U}_\pm$. This results in an exchange Hamiltonian $\hat{H}_s$ whose eigenvalues differ from those of $\hat{H}_{\rm AF}$, Eq.~(\ref{HAF}). In order to quantify  this second and more interesting effect of the SOI, and to disentangle it from the effects of a mere canting of the participating spins, we bring the exchange Hamiltonian $\hat{H}_s$ into a standard form through local spin rotations. To this end, we first rewrite $\hat{H}_s$, Eq.~(\ref{HS2}), as $\hat{H}_s/J= \hat{\boldsymbol{\sigma}}_1\tensor{h} \hat{\boldsymbol{\sigma}}_2$, in terms of a real-valued $3\times 3$ tensor $\tensor{h}$. Through a singular-value decomposition (SVD), we then bring $\tensor{h}$ into the form $\tensor{h}=\hat{R}\tensor{D}\hat{T}^T$, with  rotation matrices $\hat{R},\hat{T}\in{\rm SO(3)}$ and a real-valued diagonal matrix $\tensor{D}$. The diagonal matrices $\tensor{D}$ are the desired standardized representation of  exchange Hamiltonians. We find for our Hamiltonian (\ref{HS2}): $D_{xx}=1$, $D_{yy}=1$, and $D_{zz}=u_+ u_- + \boldsymbol{u}_+ \cdot \boldsymbol{u}_-$. The sign of $D_{zz}$ determines whether the exchange interaction in the rotated spin coordinates has antiferromagnetic ($D_{zz}>0$) or ferromagnetic  ($D_{zz}<0$) character:  Rotations by $\pi$ around the $z$ axis can be used to flip the signs of  $D_{xx}$ and $D_{yy}$, such that all entries of $\tensor{D}$ have the same sign as $D_{zz}$.

 Note that our exchange Hamiltonian (\ref{HS}) always has a pair of degenerate singular values $|D_{jj}|$, for any $\hat{U}_+$ and $\hat{U}_-$, realizing an XXZ model in rotated spin coordinates.  This  general statement is easiest understood by expressing the square of the Hamiltonian (\ref{HS}),
  \begin{equation}
4\hat{H}^2_s/J^2=(\hat{U}_+^\dagger\hat{U}_-\otimes\hat{U}_-\hat{U}_+^\dagger+\hat{U}_-^\dagger\hat{U}_+\otimes\hat{U}_+\hat{U}_-^\dagger)+2\, ,
\label{HSsquare}
\end{equation}
in a basis where the transformations $\hat{U}_+^\dagger\hat{U}_-$ acting on the left spin  and  $\hat{U}_-\hat{U}_+^\dagger$ acting on the right spin are diagonal. The only  spin operator occurring in the expression for $\hat{H}_s^2$ is then $\hat{\sigma}_{1z} \hat{\sigma}_{2z}$. Evidently, $\hat{H}_s^2$ is severely constrained: it has an Ising form in proper coordinates. This allows us to draw conclusions about $\hat{H}_s$ itself, after writing it in the general form $\hat{H}_s/J=\hat{\boldsymbol{\sigma}}_1\tensor{h} \hat{\boldsymbol{\sigma}}_2+c$, where we restored the constant $c$ that was omitted in Eq.~(\ref{HS2}). In the spin basis where $\tensor{h}$ is diagonal also the tensor representing $\hat{H}_s^2$ is diagonal and, according to Eq.~(\ref{HSsquare}), only one of its elements is nonzero. Straightforward algebra then shows that this implies $D^2_{xx}=D^2_{yy}$, so that $|D_{xx}|=|D_{yy}|$ (up to a permutation between three cartesian axes).


We are now ready to compare the  exchange Hamiltonian $\hat{H}_s^{(0)}$, Eq.~(\ref{Hs0}), obtained in  Refs.~\cite{gangadharaiahPRL08,kavokinPRB01,stepanenkoPRB03,imamuraPRB04} to our $\hat{H}_s$, Eq.~(\ref{HS2}). Representing  $\hat{H}^{(0)}_s$ by a corresponding tensor $\tensor{h}^{(0)}$, we find that also  $\tensor{h}^{(0)}$ has at least one pair of identical singular values. $\hat{H}^{(0)}_s$  thus realizes an XXZ model in a rotated spin basis, just as our  $\hat{H}_s$, Eq.~(\ref{HS2}). Eq.~(\ref{HS2}) has been derived in the strongly interacting limit under the assumption that only a single pair of (time-reversed) exchange paths contributes.  For the case  that the inverted repulsion potential has more than one  saddle point, one finds that the exchange Hamiltonian after a SVD will in general take the form of an XYZ model. The same holds true if  the orbital dynamics in the $x$ direction allows   particle exchanges at a variety of locations   $x_0$.  The resulting   exchange Hamiltonian $\hat{H}_s\to\hat{W}^\dagger(x_0)\hat{H}_s\hat{W}(x_0)$ averaged over $x_0$ may then also realize a  Heisenberg exchange with  two independent anisotropy axes, an XYZ model.

Let us turn briefly to electronic systems that are effectively 1D through the confining potential $V(y)$. In the absence of SOI, the generic antiferromagnetic  exchange Hamiltonian $\hat{H}_{AF}$  there results in the expected Luttinger-liquid behavior at low energies \cite{matveevPRL07}. Deviations from $\hat{H}_{AF}$, however,  can have profound consequences.  In Refs.~\cite{zvonarevPRL07}, for example,   spin-charge separation described by a new universality class has been found for a 1D Bose gas with ferromagnetic   Heisenberg exchange. One may thus expect important implications of our Eq.~(\ref{HS2}) for strongly interacting quantum wires. It was shown in Refs.~\cite{klironomosPRB05} that the effects of the coupling of the exchanging electron pair to surrounding electrons in 1D may be systematically included in the instanton approach, producing only small corrections in typical limits.  In a  uniform wire without SOI, the two-electron exchange Hamiltonian then carries over to the many-electron system, $\hat{H}_s^{\rm wire}=\hat{H}_s$. With SOI, however, that is no longer true, as the Hamiltonian $\hat{H}_s^{\rm wire}$ acquires a position dependence through our gauge transformation (\ref{g1D}): $\hat{H}_s^{\rm wire}=\hat{W}^\dagger\hat{H}_s\hat{W}$. The pairwise exchange interaction thus becomes  $x$-dependent  (unless $\hat{H}_s$ is invariant under $\hat{W}$), which couples the orbital motion to the spin dynamics. We furthermore remark that even if $\hat{W}$ has no effect on   $\hat{H}_s^{\rm wire}$, one  generally cannot bring the exchange Hamiltonians into a diagonal  form for all pairs of neighboring spins simultaneously by a proper choice of spin bases. The rotations in the above SVD decomposition are in general  incompatible for consecutive spin pairs.

In conclusion, we have analyzed spin exchange in effectively 1D interacting electron systems at low density. In a two-electron system, the resulting exchange Hamiltonian can be brought into the form of an anisotropic Heisenberg  model.  The model  can have both antiferromagnetic and ferromagnetic character. We have discussed general conditions for the degree of anisotropy of the resulting exchange. Both XXZ and XYZ models may be realized. Our results may have profound implications for interacting many-electron systems in one dimension.

We are grateful to Shimul Akhanjee and Arne Brataas for stimulating discussions. This work was supported in part by the Alfred~P. Sloan Foundation (YT).

\end{document}